\documentclass[12pt]{article}

\usepackage{scicite}

\usepackage{times}
\usepackage{graphicx}

\newenvironment{sciabstract}{%
\begin{quote} \bf}
{\end{quote}}

\newcounter{lastnote}

\title{Flexible Traversal Through Diverse Brain States Underlies Executive Function in Normative Neurodevelopment}

\author
{John D. Medaglia$^{1}$, Theodore D. Satterthwaite$^{2}$, Tyler M. Moore$^{2}$, Kosha Ruparel$^{2}$,\\ Ruben C. Gur$^{2}$, Raquel E. Gur$^{2}$, Danielle S. Bassett$^{3,4,\ast}$\\
\\
	\normalsize{$^{1}$Moss Rehabilitation Research Institute}\\
	\normalsize{Elkins Park, PA 19027}\\
	\normalsize{$^{2}$Brain Behavior Laboratory, Department of Psychiatry, University of Pennsylvania}\\
	\normalsize{Philadelphia, PA, 19104, USA}\\
	\normalsize{$^{3}$Department of Bioengineering, University of Pennsylvania}\\
	\normalsize{Philadelphia, PA, 19104, USA}\\
	\normalsize{$^{4}$Department of Electrical \& Systems Engineering, University of Pennsylvania}\\
	\normalsize{Philadelphia, PA, 19104, USA}\\
	\\
	\normalsize{$^\ast$ To whom correspondence should be addressed; E-mail:  dsb@seas.upenn.edu.}
}

\date{}

\begin{document}


\baselineskip24pt


\maketitle

\newpage

\begin{sciabstract}
     Adolescence is marked by rapid development of executive function. Mounting evidence suggests that executive function in adults may be driven by dynamic control of neurophysiological processes. Yet, how these dynamics evolve over adolescence and contribute to cognitive development is unknown. Using a novel dynamic graph approach in which each moment in time is a node and the similarity in brain states at two different times is an edge, we identify two primary brain states reminiscent of intrinsic and task-evoked systems. We demonstrate that time spent in these two states increases over development, as does the flexibility with which the brain switches between them. Increasing time spent in primary states and flexibility among states relates to increased executive performance over adolescence. Indeed, flexibility is increasingly advantageous for performance toward early adulthood. These findings demonstrate that brain state dynamics underlie the development of executive function during the critical period of adolescence.
\end{sciabstract}


\newpage

As the human brain matures, the structural and functional relationships among brain areas reorganize \cite{supekar2010development}. Long distance links strengthen\cite{fair2009functional} between the highly connected, influential regions and regions located on the outskirts of the network \cite{hwang2013development}. Such reconfiguration is critical for forming the increasingly interactive, integrated circuit \cite{Cao2014,Betzel2014,Gu2015} thought to drive adaptive executive functions in adulthood \cite{Heinzle2012,Ekman2012}. Yet, how these changes relate to evolving executive performance in an adolescent \cite{Anderson2002} is unknown.  Emerging evidence in adults highlights the flexibility of local (frontal) systems as a predictor of individual differences in executive ability \cite{Cole2013,Braun2015}, as well as the flexibility of global (whole-brain) networks as a predictor of individual differences in adaptive behaviors \cite{Bassett2015}.

However, theoretical tensions have hampered an integrated understanding of developing executive capabilities. In a single brain region, neural processes produce increased signal complexity during development \cite{Vakorin2011}, potentially enabling brain regions to express a larger repertoire of microstates. Conversely, the whole brain displays increased integration between widely distributed neuronal populations, potentially enabling the brain to express a more focal repertoire of  microstates \cite{Khanna2015,Vakorin2011}. Resolving these conflicting predictions requires a framework that examines the stable global features of brain dynamics while remaining sensitive to local variation \cite{TurkBrowne2013}.

Here we propose an integrative framework that treats the brain as a dynamical system with both local and global contributions to executive function that evolve over development to enable individual differences in cognition. Specifically, we posit two dynamic drivers of executive function: (i) traversal of many different brain states (which we define below), and (ii) a preference for a few states \cite{Khanna2015}. These features support both flexible transitions between cognitive processes \cite{Khanna2015,Shanahan2010} and stable task performance. We hypothesize that improvements in executive performance in development will rely on the brain's ability to transition between states and the ability to return to common states. Finally, we predict that inter-individual variability in the time spent in common states and the flexible state transitions predicts individual differences in executive function.

\begin{figure}
	\centerline{\includegraphics[width=7in]{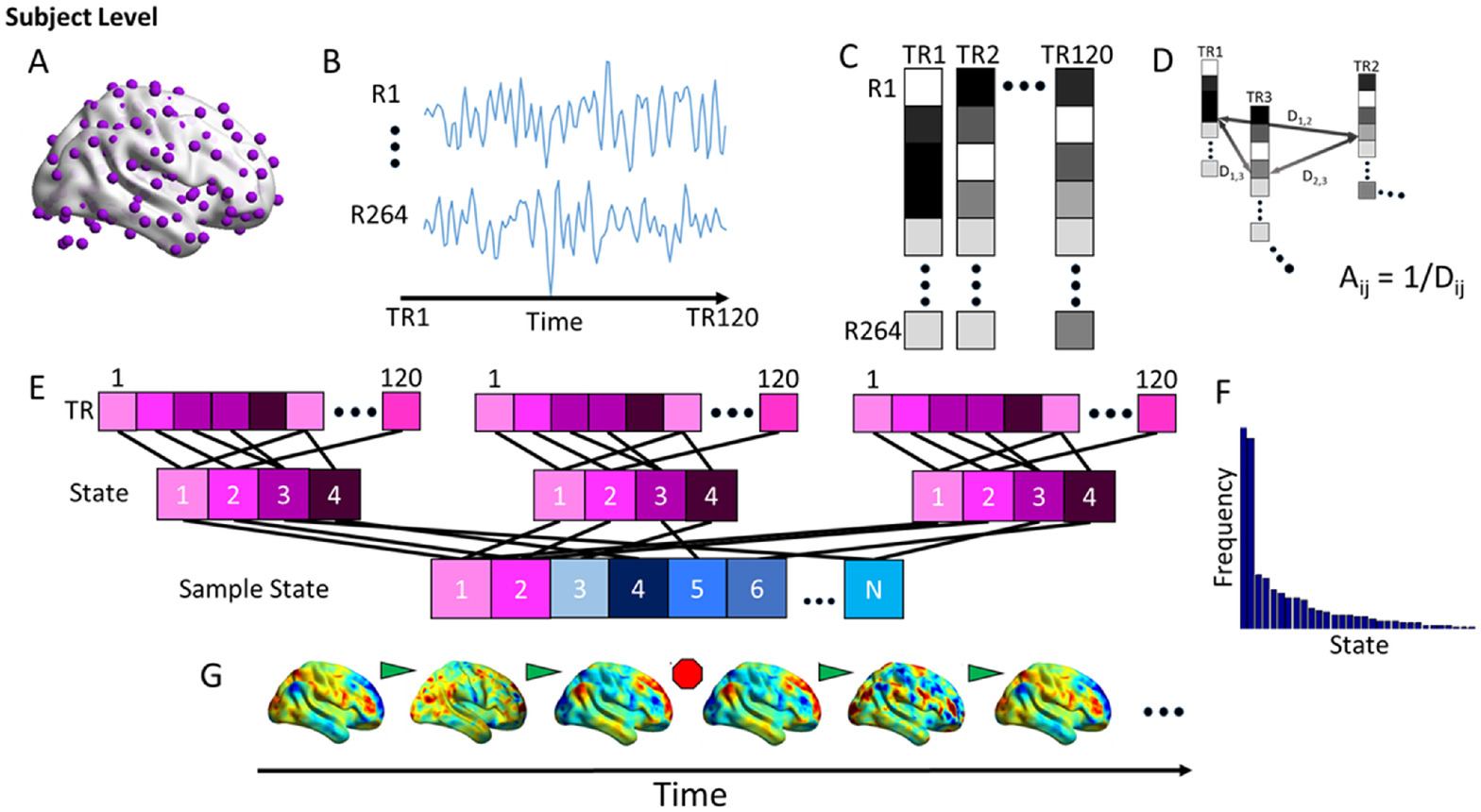}}
	\caption{\textbf{Schematic of Methods.}  (A) In each subject, we define 264 cortical and subcortical regions of interest (ROIs) \cite{Power2011}. (B) We extract mean resting BOLD time series from each ROI. (C) We define the momentary (one volume at time resolution; TR) pattern of BOLD magnitude across ROIs as a brain state, which we represent as an $N \times 1$ dimensional vector. (D) We then compute the Euclidean distance between every pair of state vectors. We summarize pairwise distances between states in a $T \times T$ adjacency matrix, where $T=120$ TRs. (E) For each subject, we distill approximately 4 common states by applying a network-based clustering technique to the adjacency matrix. We then map correspondence between subject-level states by performing a group-level clustering procedure. (F) We study the frequency distributions of states and observe the presence of two common states as well as a heavy tail of less common states. After defining these states, we estimate the time spent in a state and the flexible shifting between states (G).
		\label{fig1}}
\end{figure}

To address these hypotheses, we introduce a dynamic graph approach to examine brain states and flexible transitions between states in the Philadelphia Neurodevelopmental Cohort (PNC), which includes 780 typically developing youth from the ages of 8 to 21 (see Fig.~\ref{fig1}, Supplement).  Specifically, we define the momentary (single TR) pattern of fMRI BOLD activity across $N=264$ functionally-defined areas as a brain state. Next, we map a unit of time to a network node, and we map similarities in brain states at two different times to a network edge. In this time-by-time network, we can use network-based clustering algorithms in individual subjects to define common brain states independent of their temporal order.

\begin{figure}
\centerline{\includegraphics[width=3.5in]{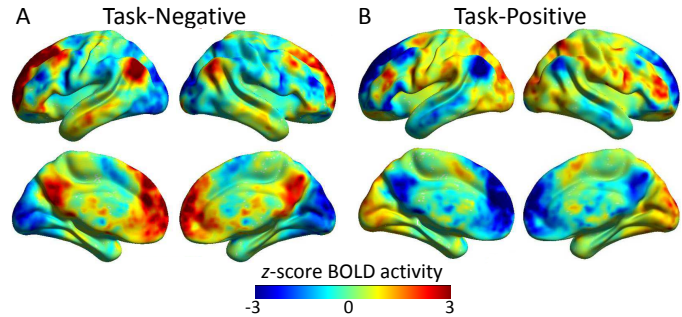}}
	\caption{\textbf{Two Primary States.} Two states were consistently observed over age. One state demonstrated high activity in the default mode network, and the other across so-called \emph{task-positive} regions from the dorsal attention, cingulo-opercular, and visual systems. \label{fig2}}
\end{figure}

From this state ensemble, we quantify the extent to which different individuals display the same or different states, and how this convergence varies as a function of development. In ten evenly spaced age bins, we define group-level states by applying the same clustering techniques to a matrix of distances calculated between all states observed across individuals. In the youngest age bin, we observe two states that frequently occur across all subjects (Fig.~\ref{fig2}). We refer to these states as \emph{primary} states, and note that the first primary state displays high activity in brain regions traditionally observed to be active at rest (hereafter \emph{task-negative}), and the second primary state displays high activity in regions traditionally observed to be active during cognitive tasks (\emph{task-positive}). These two common states that we observe in the youngest age bin show similar patterns of BOLD magnitude to the two common states that we observe in all other age bins: the mean correlation in the regional patterns of BOLD magnitude among task-negative states extracted from each age group is $r=0.90$ ($SD = 0.08$, $df=262$) and that among task-positive states is $r=0.92$ ($SD = 0.05$, $df=262$). Thus, the two primary states are robustly observed over neurodevelopment.

The presence of these two \emph{primary} states complements an extensive literature describing the default mode and task-positive systems as large-scale opponent processes driven by spontaneous local neuronal dynamics \cite{Deco2011}. Yet, they only comprise approximately $42$\% ($SD=26.8$) of all TRs. The remaining $58\%$ of TRs includes states that each occur relatively infrequently, and we refer to these as \emph{secondary states}. The presence of both primary and secondary states suggest that youths tend to predominantly transition between task-positive and task-negative states interspersed with a rich landscape of other states.

\begin{figure}
	\centerline{\includegraphics[width=3.5in]{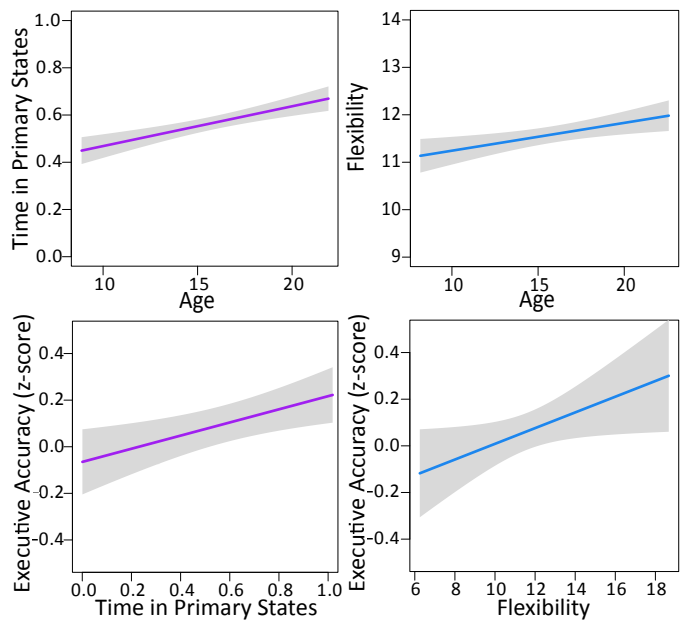}}
	\caption{\textbf{Dynamic Development and Executive Performance.} (A) Time in two primary states and flexibility both increase over development. Adolescents were noted to shift an average of 11.5 times relative to number of observed states. (B) In addition, time in two primary states and state flexibility are positively related to executive performance beyond the primary effect of age.  Both trends are robust to motion and sex. Gray envelopes represent 95\% confidence interval for the line of best fit. \label{fig3}}
\end{figure}

To examine the extent to which brain states vary across adolescence, we define two summary statistics: the time spent in primary states and the flexibility between states. We estimate the time spent in the two primary states as the number of volumes (TRs) assigned to those states. We observe that the time spent in the two primary states increases with age (controlling for motion and sex in this model and henceforth; $F(4,775) = 14.15$, $p <0.001$) (Fig.~\ref{fig3}A). Children around 8 years of age spend approximately 32.7\% of TRs in the primary states, while young adults around 21 years of age spend approximately 61.9\%. This result indicates that these two robust states exist early in development and display an increasing presence over adolescence. To understand how the brain transitions between states (both primary and secondary), we define the state flexibility ($F$) of a subject to be the number of state transitions ($T$) that occur relative to the number of states ($S$) observed ($F = T/S$). State flexibility increases over age ($F(4,775) = 7.78, p < 0.001$). Collectively, these results demonstrate that as the brain develops, it spends more time in primary states, while -- somewhat surprisingly -- also becoming increasingly flexible: the Pearson correlation between the two variables is $r = 0.16$ ($p = 3.5 \times 10^{-6}$) corrected for age, motion, and sex. Conceptually, these findings suggest that the brain may offer a careful balance between two seemingly competitive processes: the consolidation of the brain's dynamic repertoire toward adulthood \cite{Khanna2015} and the growth of flexible dynamics potentially enabling functional diversity.

Finally, we ask whether variation in state occupancy across individuals predicts executive performance beyond that explained by age: executive accuracy increases with age ($F(4,775) = 29.52$, $p <0.001$). We observe that individual differences in the time spent in the two primary states and state flexibility are positively related to executive accuracy controlling for age, sex, and motion: $F(4,775) = 6.13$, $p < 0.001$ and $F(4,775) = 5.24$, $p < 0.001$, respectively. Using a single model, the specific effect of time in primary states remains significant ($t(776) = 2.48$, $p = 0.013$) but the effect of state flexibility does not ($t(776) = 1.63$, $p = 0.103$). Critically, these results support the notion that the greater executive performance characteristic of adulthood is supported by an increasing refinement of brain state dynamics characterized by greater time spent in primary states and greater flexibility of state transitions. These results nuance previous theories that have focused on the transient nature of functional interactions \cite{Cole2013}, by suggesting that such transient processes must also be accompanied by stable state maintenance \cite{Braun2015,Bassett2015}. Indeed, our data suggest that momentary maintenance and transition in global brain states together provide a crucial underpinning for executive performance.

\begin{figure}
	\centerline{\includegraphics[width=3.5in]{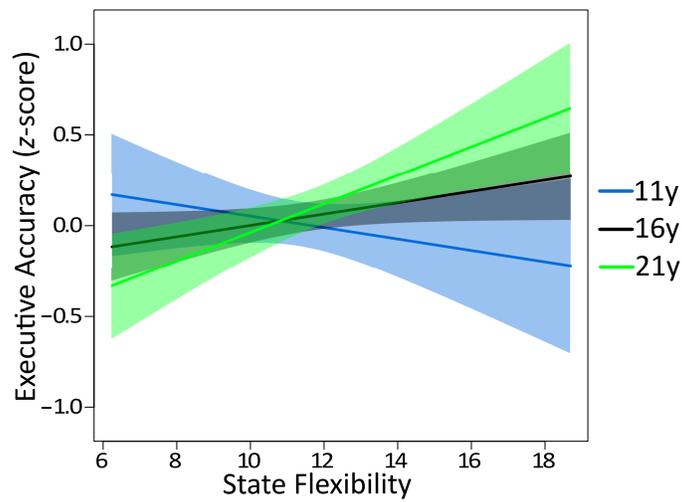}}
	\caption{\textbf{Interaction Between State Flexibility and Age.} State flexibility is increasingly predictive of executive performance as age increases. Early in development, increased flexibility is negatively associated with executive accuracy. By early adulthood, increased flexibility is positively associated with executive accuracy in the context of increased time spent in primary states, demonstrating a tightening organization of the brain's dominant states and accelerating state transitions underlying executive performance. Shaded envelopes represent 95\% confidence interval for the lines of best fit. \label{fig4}}
\end{figure}

The global relationship between brain state dynamics and individual differences in executive performance does not address the question of whether brain state dynamics differentially drive cognitive variation in younger children \emph{versus} older children. Such an inversion has previously been observed in the relationship between cortical thinning and age \cite{Shaw2006}. To address this question, we consider interactions between age and brain state dynamics on executive accuracy. The relationship between executive performance and the time spent in primary states shows no significant interaction with age ($t(776) = 1.10$, $p = 0.270$). In contrast, the relationship between executive performance and state flexibility does show a significant interaction with age ($t(776) = 2.06$, $p = 0.039$): greater state flexibility was related to poorer executive performance in children, and better executive performance in young adulthood (Fig.~\ref{fig4}). We speculate that the brain's exploration of many cognitive states in childhood is disadvantageous for cognitive control, but advantageous for learning specifically and behavioral adaptation more generally \cite{Gopnik2015,Munakata2015,Thompson-Schill2009}. Over development, this mode of cognitive exploration turns to a mode of exploitation, were a streamlined set of cognitive modes are under tight executive control \cite{Cohen2007}.

Collectively, these results uncover a set of common and uncommon states across individuals, whose temporal presence and flexible nature underlie the evolving executive abilities characteristic of normative neurodevelopment. The common states -- which prominently involve (i) the default mode network and (ii) task control and attention systems -- exist at all ages examined and the time spent in these two primary states increases in frequency over development. This characteristic of brain state dynamics bears crucial implications for developmental studies of functional brain connectivity \cite{Supekar2009,Hwang2013,fair2009functional,Cao2014,Betzel2014,Gu2015}. Observed changes in functional network structure derived from resting BOLD time series may be a direct result of different -- and dynamic -- combinations of a finite set of brain states, the most common of which already comprise approximately one third of brain states by age 8 and two thirds of brain states by early adulthood. Indeed, these findings place a spotlight on the critical need to understand the relationship between brain activity and brain connectivity as the two approaches continue to offer differential insights into cognitive function \cite{Bassett2012,Bassett2015,Medaglia2015} but are physically interdependent and only partially accounted for by structural network architecture \cite{Hansen2015}.

Our results are particularly intriguing in light of the theory that the brain is a metastable system. In physics, metastable systems tend to spend significant time in states other than the least energy state \cite{Shanahan2010}. Although not analytically addressed in our work (we have no definition of state energy), our results do offer conceptually similar notions: the brain traverses many different states (a broad dynamic repertoire) while having a preference for a few primary states (a narrow dynamic repertoire). This careful balance underlies individual differences in executive function. Our approach complements prior efforts in studying putative metastable dynamics in the brain, including variability in phase coherence \cite{Hellyer2014} and attractors \cite{Deco2012}. In contrast, we define a brain state as the pattern of regional BOLD activation at a single brain image, distill common states in a person and across persons, and define summary statistics to quantify state flexibility and persistence. Individual differences in the types, frequency, and transitions among observed states may provide crucial information regarding the dynamic underpinnings of executive cognition in the human brain.

See the Supplement for further methodological details, supporting results, and discussion.

In conclusion, we demonstrate the compelling utility of network approaches beyond traditional functional and structural connectivity by exploring brain states and the dynamics of their transitions. Two frequently observed brain states exist early in development and become more prominent in young adulthood. These primary states are complemented by a rich set of less frequent states that differ across individuals. Flexible transitions between these states support developing executive function. These findings motivate future work to determine whether brain state dynamics provide a basis for understanding the development of the functional connectome, the emergence of cognition, and alteration of these biomarkers in psychopathology.

\clearpage
\newpage

\bibliography{JDMReferences,bibfile}

\bibliographystyle{Science}



\end{document}